\documentclass[9pt,twocolumn,twoside]{opticajnl}
\journal{opticajournal} 

\setboolean{shortarticle}{false}

\usepackage{lineno}

\newcommand{\neff}{n_\text{eff}}
\newcommand{\be}{\begin{equation}}
\newcommand{\ee}{\end{equation}}
\newcommand{\dk}{\delta k_0}
\newcommand{\pd}{\partial}
\newcommand{\mum}{~\mu \text{m}}

\title{Fundamental Limits on Fiber-Based Electron Acceleration --- and How to Overcome Them}

\author[1,2]{Aku Antikainen}
\author[2]{Siddharth Ramachandran}

\affil[1]{Institute of Optics, University of Rochester, Rochester, NY, USA}
\affil[2]{Photonics Center, Boston University, Boston, MA, USA}

\affil[*]{aku.antikainen@rochester.edu}

\begin{abstract}
To accelerate ultra-relativistic charged particles, such as electrons, using an electromagnetic pulse along a hollow-core waveguide, the pulse needs to have a longitudinal electric field component and a phase velocity of $c$, the speed of light in vacuum. We derive an approximate closed-form expression for the wavelength at which the phase velocity of the TM$_{01}$ mode in a metal-clad hollow-core fiber with a dielectric layer is $c$. The expression is then used to derive conditions for material dispersion required of the dielectric in order to simultaneously have $c$ phase and group velocity. It is shown that the dispersion would need to be so heavily anomalous that the losses in the anomalously dispersive regime would render such a particle accelerator useless. We then propose the utilization of gain in the form of two spectral peaks in the dielectric to circumvent the otherwise fundamental limits and allow for TM$_{01}$ pulses with $c$ phase and group velocity and thus arbitrary length-scaling of fiber-based electron accelerators. In theory, the group velocity dispersion could also be made zero with further gain-assisted dispersion engineering, allowing for the co-propagation of dispersionless electromagnetic pulses with relativistic particles.
\end{abstract}

\setboolean{displaycopyright}{false} 

\begin{document}

\maketitle

\section{Introduction}

Free electron lasers, particle colliders, electron microscopy, medical imaging and therapy, and X-ray science would all benefit tremendously from compact, cost-effective electron accelerators. Conventional radio frequency accelerators require high average powers and long pulses, and are limited to accelerating gradients in the $10$ MeV$/$m range. Near-infrared (NIR) sources based on chirped pulse amplification can achieve accelerating gradients that are larger by several orders of magnitude, but they inherently suffer from small electron bunch size and other issues related to the short wavelengths and the associated sensitivity to timing. In a lot of respects, THz radiation offers a promising compromise, combining high accelerating gradients, longer wavelengths compared to NIR, and short pulses, leading to better energy-efficiency \cite{Wang2022, Turnar2022, Fisher2022}. In particular, TM-polarized few-cycle THz pulses in hollow-core waveguides have been demonstrated to be a viable platform for relativistic electron acceleration \cite{Nanni2015, Zhang2022}. The use of TM-modes for electron acceleration is based on these modes having an electric field component in the direction of propagation \cite{Varin2002}.

Hollow-core waveguides have the benefit of preventing diffraction of the accelerating pulse, leading to larger acceleration gradients due to the confinement of the beam. Even though stable propagation of TM-modes over long distances in optical fibers has been demonstrated \cite{Ramachandran2009}, the electron acceleration studies of \cite{Nanni2015, Zhang2022} limited themselves to modest waveguide lengths of a couple of centimeters. The reason for this is that the group velocity of the THz pulse is vastly different from the velocity of the relativistic electrons, causing walk-off between the pulse and the electrons over the centimeter length scale. Relativistic electron acceleration in hollow-core fibers will not benefit from longer fibers unless both the phase and group velocity of the pulse performing the acceleration can simultaneously be made equal to $c$, the speed of light in vacuum.

Here we show that simultaneously having $c$ phase and group velocity for a TM-mode in a hollow core fiber requires the use of materials with heavy anomalous dispersion as part of the fiber design, and an analytical dispersion condition is derived. Such heavy anomalous dispersion can (due to Kramers-Kronig relations) only occur in a spectral region where losses increase or where \emph{gain decreases}. We thus propose using a gain element as part of the fiber to induce the necessary anomalous dispersion, allowing for a fiber design with $c$ phase and group velocity for a TM mode. Such a fiber makes it possible to arbitrarily scale the length of the electron accelerator and hence the use of temporally longer, less intense accelerator pulses that also experience gain upon propagation. The dimensions of the fiber design proposed here correspond to an operating wavelength in the THz regime, but the results are wavelength-agnostic due to the scale-invarience of Maxwell's equations and apply to all operating wavelengths, THz and NIR alike.

\section{Linear Hollow-Core Accelerators}

\begin{figure}[h!]
\centering\includegraphics[width=\columnwidth]{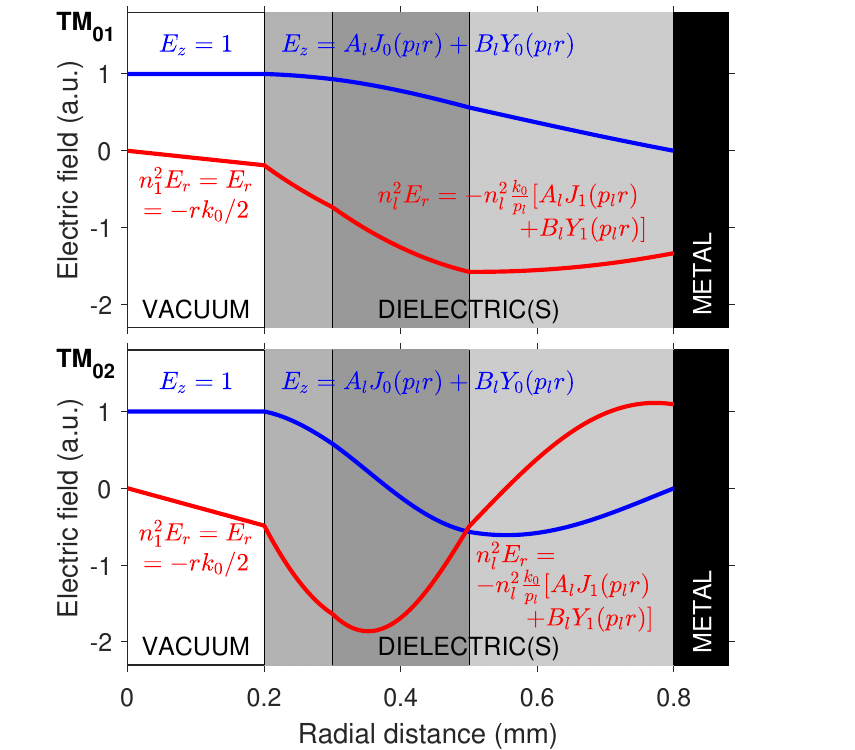}
\caption{A hollow-core fiber consisting of three dielectric layers and the metal cladding. The electric field profiles of the TM$_{01}$ (top) and TM$_{02}$ (bottom) modes at the wavelength for which the phase velocity is equal to $c$ are shown. The plots show the two quantities related to the electric field that are continuous across the dielectric interfaces: the longitudinal electric field component $E_z$ (blue lines), and the radial field component times the square of the local refractive index $n^2 E_r$ (red lines). The functional forms of the field profiles in each layer $l$ have been indicated.}
\label{fig1}
\end{figure}

Any waveguide designed for electron acceleration needs to have a vacuum core. A hollow-core fiber with nothing but dielectric layers surrounding the vacuum core has the property that the phase velocity of every mode is always faster than the speed of light \cite{Yeh1978}, creating an obvious issue for acceleration of massive particles, such as electrons. Note that "mode" here means the Bragg modes that have significant optical power in the vacuum core, not the (non-lossy) cladding modes confined in the dielectric layers. By enclosing the waveguide in an outer metal cladding, the phase velocity at a specific wavelength for a certain TM mode can be made to coincide with the speed of light in vacuum. This is the operating wavelength of the relativistic electron accelerator, in which the electrons can be assumed to move at the speed of light. An illustration of the structure of such a (rotationally symmetric, circular) waveguide is shown in Fig.~\ref{fig1}.

The operating wavelength $\lambda_\text{op}$ of the fiber for any given TM mode can be determined from the condition that the phase velocity $v_p(\lambda_\text{op}) = c$ at that wavelength. A TM mode has no longitudinal magnetic field component, and every field component of the mode can be directly determined from the longitudinal electric field component $E_z$. The functional form of $E_z$ in each layer $l$ is known (from Maxwell's equations) to be a linear combination of the two Bessel functions:
\be
E_z^l(r) = A_l J_0 (p_l r) + B_l Y_0 (p_l r) ,
\label{eq:Ez}
\ee
where $p_l = k_0 \sqrt{n_l^2 - \neff^2}$, $k_0 = 2 \pi / \lambda_0$ is the vacuum wavenumber, $n_l$ is the refractive index of the dielectric layer $l$, and $\neff$ is the effective index of the mode ($\neff = 1$ for $v_p = c$). The mode is determined by the following boundary conditions: 1. The energy of the mode must be finite (i.e. there can be no field singularities), 2. The longitudinal electric field $E_z(r)$ must be continuous from one dielectric layer to another, 3. The radial field times the local refractive index squared, $n(r)^2 E_r(r)$, must be continuous from one dielectric layer to another, and 4. the longitudinal field $E_z(r)$ has to be zero at the metal, as metals can be well approximated as perfect conductors at NIR and THz wavelengths \cite{Kirley2015}.

In the vacuum core we have $B_1 = 0$ to avoid a singularity of $Y_0(p_1 r)$ at $r = 0$, and the longitudinal electric field of any TM mode is given by $A_1 J_0(p_1 r)$, where $p_1 = k_0 \sqrt{1 - \neff^2}$. The field distributions of each mode are unique up to a multiplicative factor, so without loss of generality we can set $A_1 = 1$, and the longitudinal field becomes $E_z(r) = J_0(p_1 r)$ in the core. At the wavelength(s) for which the phase velocity is equal to $c$, we have $\neff = 1$ and therefore $p_1 = 0$, and the field simplifies to $E_z(r) = 1$ in the core, as also shown in Fig.~\ref{fig1}. In any further dielectric layers $l$, the transverse wavenumber $p_l$ is not identically zero (unless there are more vacuum layers in addition to the vacuum core), and the field needs to be described by the general form, Eq.~(\ref{eq:Ez}).

\section{Waveguide Dispersion of Metal-Clad Hollow-Core Fibers}

In this section, we assume the dielectrics in the fiber are dispersionless, and focus on waveguide dispersion. Consider a general, circularly symmetric, metal-clad linear accelerator consisting of the vacuum core and dispersionless dielectric layers, such as a the one shown in Fig.~\ref{fig1}. Let $k_0 = 2 \pi / \lambda_0$ be the vacuum wavenumber corresponding to an operating wavelength $\lambda_0$ and $\dk$ a small ($|\delta| << 1$) perturbation. In order for the group velocity to be $c$ at the operating wavelength, the effective index of the TM mode must not change in first order in the presence of the perturbation $\dk$. Thus, the field $E_z$ is to remain unity at the vacuum-dielectric interface, and similarly $E_z$ must still be zero at the metal cladding, both to first order in $\dk$. The endpoints of the mode field $E_z$ in the dielectrics are thus anchored to unity on one side and zero on the other.

First we write the perturbed wavenumber as $k_0 + \dk = (1 + \delta) k_0$. In the absence of material dispersion, all the perturbed transverse wavenumbers $p_l'$ simply become $p_l' = (1 + \delta) p_l$ to first order, since the effective index is to remain at 1. The functional form of the perturbed fields $E_z'$ and $E_r'$ in the core is now
\be
	E_z'(r) = 1 ~\text{ and }~ E_r'(r) = -\frac{(1 + \delta) r k_0}{2} . \label{eq:Eprimecore} 
\ee
In the dielectric layers, the fields (or, strictly speaking, the radial dependencies of the field components) are given by
\begin{align}
	{E_z^l}'(r) &= A_l' J_0 \left[(1 + \delta) k_0 r \sqrt{n_l^2 - 1} \right] \label{eq:Ezprimediel} \\
	            &+ B_l' Y_0 \left[ (1 + \delta) k_0 r \sqrt{n_l^2 - 1} \right] \nonumber \\
	{E_r^l}'(r) &= - \frac{1}{\sqrt{n_l^2 - 1}} \left\{ A_l' J_1 \left[ (1 + \delta) k_0 r \sqrt{n_l^2 - 1} \right] \right. \label{eq:Erprimediel} \\
	 &+ \left. B_l' Y_1 \left[ (1 + \delta) k_0 r \sqrt{n_l^2 - 1} \right]   \right\} ,  \nonumber
\end{align}
where the primes refer to the perturbed fields, and the Bessel function coefficients $A_l$ and $B_l$ are also primed because they might have changed to keep the modal boundary conditions satisfied. The transverse wavenumbers have been expanded as $p_l = k_0 \sqrt{n_l^2 - 1}$ such that Eqs. (\ref{eq:Eprimecore} - \ref{eq:Erprimediel}) readily show the perturbation factor $1 + \delta$ only appearing in front of products of the radial distance $r$ and the unperturbed wavenumber $k_0$. Even though we originally introduced the perturbation to the wavenumber $k_0$, we can now group the perturbation factor together with the radius $r$ instead, since the perturbation factor $1 + \delta$ only appears in front of products of the form $k_0 r$. This means that the requirement that the group index is unity, i.e. that the modal effective index could be kept at unity to first-order in $\delta$ when the wavenumber is perturbed from $k_0$ to $(1 + \delta)k_0$, is equivalent to the operating wavelength staying at $\lambda_0$ to first-order in $\delta$ when the \emph{radii} of the fiber are perturbed from $a_l$ to $(1 + \delta) a_l$. However, the scale invariance of Maxwell's equations shows the obvious contradiction here. If the operating wavelength of a structure described by material indices $n_1 = 1, n_2, \ldots$ and radii 
$a_1, a_2, \ldots$ is $\lambda_0$, then the operating wavelength of a structure described by the same material indices $n_1 = 1, n_2, \ldots$ but scaled radii $(1 + \delta) a_1, (1 + \delta) a_2, \ldots$ is $(1 + \delta) \lambda_0$. In other words, the operating wavelength always changes to first order in $\delta$ upon scaling the radii by $1 + \delta$, which then translates into the impossibility of any TM-mode having unity group index for non-dispersive dielectrics.

An intuitive way to understand why the effective index needs to change when the wavelength changes is that when $n_\text{eff} = 1$, $E_z$ is anchored to unity at the vacuum-dielectric interface and to zero at the metal cladding, and changing the wavelength causes at least one of these conditions to be no longer satisfied. An analogy would be that a spring with each of its ends attached to walls cannot be compressed locally without stretching it somewhere else. If the wavelength is decreased, $k_0$ increases, and the Bessel functions would oscillate faster in $r$ in the dielectric layers. To keep the boundary conditions satisfied, the effect of the increase in $k_0$ on the transverse wavenumbers ($p_l = k_0 \sqrt{n_l^2 - \neff^2}$) would need to be compensated for by an increase in $\neff$. This applies to all such fiber designs and all radial TM mode orders, but is only true for non-dispersive materials. As the expression for the transverse wavenumbers $p_l = k_0 \sqrt{n_l^2 - \neff^2}$ shows, an increase in $k_0$ could also be compensated for by a decrease in $n_l$ instead of an increase in $\neff$, and by properly selecting the slope $d n / d k_0$, the group index could be designed to be unity at the operating wavelength. However, this means that $n_l(\lambda)$ would need to increase with $\lambda$, which corresponds to anomalous material dispersion.

Since higher radial-order TM modes are subject to the same boundary conditions as lower-order TM modes, they offer no obvious advantage over TM$_{01}$. Furthermore, since at least one of the dielectric layers would need to be anomalously dispersive, any additional non-dispersive or normally dispersive layers would only make things worse in the sense that the anomalous layer(s) would have more to compensate for (i.e. have even stronger anomalous dispersion) in order to make the group index unity. We will thus first focus on the simplest case: TM$_{01}$ in a fiber consisting of a vacuum core, a single dielectric layer, and a (perfect) metal cladding. In the next section we introduce a convenient new mathematical tool to aid in fiber design.

\section{The Sine-Taylor Method}

Determining the operating wavelengths (i.e. the wavelengths at which the effective index is unity) in a simple fiber consisting of a vacuum core surrounded by a dielectric and and a metal cladding is straightforward. The three degrees of freedom are $A_1$, which is the electric field strength in the core, and $A_2$, and $B_2$, which are the amplitudes of Bessel functions $J_0$ and $Y_0$, respectively, in the dielectric. The boundary condition equations can be written in matrix form as
\be
	\begin{bmatrix}
		1 
		& -J_0\left( p a_1 \right) 
		& -Y_0\left( p a_1 \right) 
		  \\ 
		-\frac{k_0 a_1}{2} 
		&  \frac{n^2}{\sqrt{n^2 - 1}} J_1\left( p a_1 \right) 
		&  \frac{n^2}{\sqrt{n^2 - 1}} Y_1\left( p a_1 \right) 
		  \\ 
		0                  
		&  J_0\left( p a_2 \right)
		&  Y_0\left( p a_2 \right) \\ 
	\end{bmatrix}
	\begin{bmatrix}
			A_1 \\ A_2 \\ B_2
	\end{bmatrix}
	=
		\begin{bmatrix}
			0 \\ 0 \\ 0
	\end{bmatrix} , \label{eq:matrix}
\ee
where $p = k_0 \sqrt{n^2 - 1}$, $a_1$ is the vacuum core radius, $a_2$ is the outer radius of the dielectric (which is the inner radius of the metal cladding), and $n$ is the refractive index of the dielectric, which we now allow to depend on wavelength. The first row of the matrix in Eq.~(\ref{eq:matrix}) captures the continuity of $E_z$ from the vacuum to the dielectric, the second row the continuity of $n^2 E_r$, and the third row the requirement that $E_z(a_2) = 0$. The opearating wavelengths of the fiber are those for which the determinant of the equation matrix in Eq.~(\ref{eq:matrix}) is zero. The determinant, when written in its expanded form, is a fairly lengthy expression involving products of various Bessel functions, and obviously there is no closed form formula to find its roots, even for constant $n$, and the roots have to be computed numerically.

When designing a fiber for relativistic electron acceleration, we can use approximations to get an estimate for the desired operating wavelength. The actual operating wavelength is hardly important, as it can be changed by simply scaling the transverse dimensions of the fiber. Furthermore, practical pulse sources for electron acceleration, such as THz sources, have room for tunability when it comes to their operating frequency. We can therefore simplify the modal analysis through approximations.

\begin{figure}[h!]
\centering\includegraphics[width=\columnwidth]{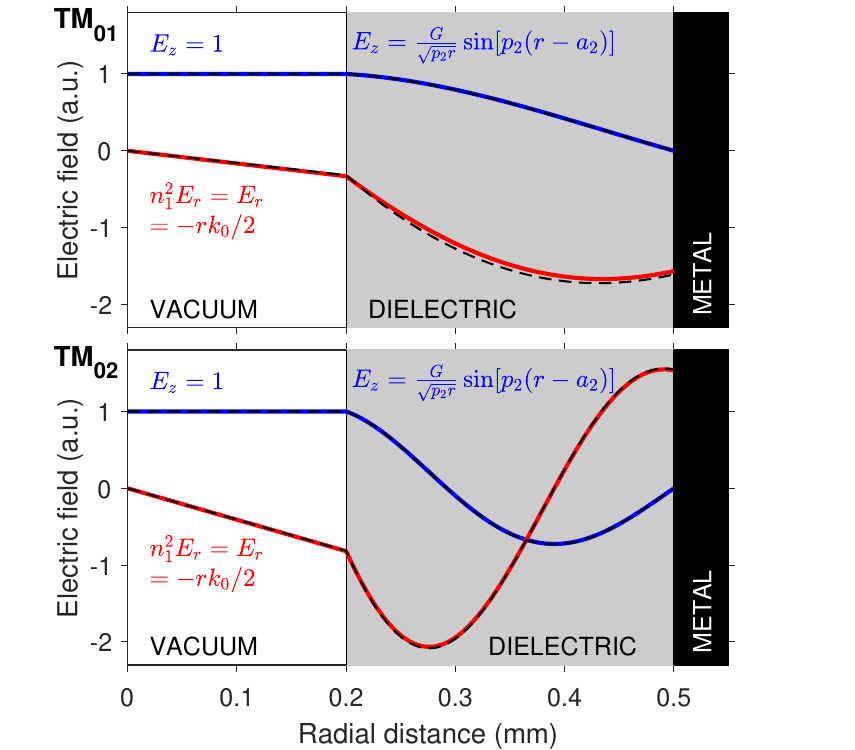}
\caption{Approximations to the electric field profiles of the TM$_{01}$ (top) and TM$_{02}$ (bottom) at their respective operating wavelengths using the sinusoid mode determination method described above. Blue lines: longitudinal electric field component $E_z$, red lines: radial field component $E_r$ times the local refractive index $n$. The black dashed lines are the real field profiles involving Bessel functions.}
\label{fig:sine1}
\end{figure}

First we note that we can avoid dealing with Bessel functions if we replace them with their asymptotic approximations:
\begin{align}
	J_\nu (z) &= \sqrt{ \frac{2}{\pi z} } \left[ \cos\left(z - \frac{\nu \pi}{2} - \frac{\pi}{4} \right) + e^{|\text{Im}(z)|} \mathcal{O} \left(|z|^{-1} \right) \right]   \label{eq:Jappr} \\
	Y_\nu (z) &= \sqrt{ \frac{2}{\pi z} } \left[ \sin\left(z - \frac{\nu \pi}{2} - \frac{\pi}{4} \right) + e^{|\text{Im}(z)|} \mathcal{O} \left(|z|^{-1} \right) \right]   \label{eq:Yappr} .
\end{align}
Equations~(\ref{eq:Jappr}) and (\ref{eq:Yappr}) can be plugged into the expression for $E_z(r)$, from which the corresponding $E_r(r)$ can be determined through differentiation with respect to $r$. These forms for the fields can be used to derive a matrix equation similar to Eq.~(\ref{eq:matrix}), the determinant of which will be simpler and free of Bessel functions, though still not admitting closed form solutions. However, before plugging the approximate expressions into the matrix equation, a further simplification can be made by noting that with the approximations the functional form of $E_z$ in the dielectric layer becomes
\begin{align}
	E_z(r) = \sqrt{ \frac{2}{\pi k_0 r \sqrt{n^2 - 1}} }  
		  &\left[A_2 \cos\left(k_0 r \sqrt{n^2 - 1} - \frac{\pi}{4} \right) \right. \\ 
		+  &\left.B_2 \sin\left(k_0 r \sqrt{n^2 - 1} - \frac{\pi}{4} \right) 
																			 	  \right] , \nonumber
\end{align}
which can be written as
\be
	E_z(r) = G \sqrt{ \frac{2}{\pi k_0 r \sqrt{n^2 - 1}} } 
		\sin \left( k_0 r \sqrt{n^2 - 1} - \varphi_0 \right) , \label{eq:sine}
\ee
where $G$ and $\varphi_0$ are constants related to $A_2$ and $B_2$. This is a mere manifestation of the fact that a linear combination of two sinusoids is a sinusoid. Now instead of choosing the constants $A_2$ and $B_2$ such that the boundary conditions are satisfied, we are dealing with $G$ and $\varphi_0$. With the form of $E_z(r)$ given in Eq.~(\ref{eq:sine}), a matrix form of the boundary condition equations is no longer possible, as one of the parameters, $\varphi_0$, is inside the sinusoid. However, from the metal cladding boundary condition we know $E_z(a_2) = 0$, and this can only be satisfied if
\be
	\varphi_0 = k_0 a_2 \sqrt{n^2 - 1} + m \pi,
\ee
where $m$ is an integer. Next we note that $\sin(x + m \pi) = \sin(x)$ when $m$ is even, and $\sin(x + m \pi) = - \sin(x)$ when $m$ is odd. The constant $G$ in Eq.~(\ref{eq:sine}) is still unknown, and we can thus set $m = 0$ and absorb the plus or minus sign in front of $\sin(x)$ as well as the factor $\sqrt{2/\pi}$ into $G$ without any loss of generality. The electric field $z$-component in the dielectric is therefore approximately
\be
	E_z(r) = \frac{G}{\sqrt{p_2 r} } 
		\sin \left[ p_2 (r - a_2) \right] \label{eq:sine2} ,
\ee
where $p_2 = k_0 \sqrt{n^2 - 1}$. The metal cladding boundary condition is automatically incorporated into this approximate form of $E_z(r)$, and the only degrees of freedom that are left are $A_1$, the value of the electric field in the core that we would eventually set to unity, and $G$, and hence the resulting matrix equation would only involve a 2x2 matrix, the determinant of which is considerably simpler. Figure~\ref{fig:sine1} shows the predictions for the TM$_{01}$ and TM$_{02}$ field profiles using the sine approximation method for a fiber with $a_1 = 200 \mum$, $a_2 = 500 \mum$, and $n = 2.1$ (the refractive index of quartz in the THz regime).

\begin{figure}[h!]
\centering\includegraphics[width=\columnwidth]{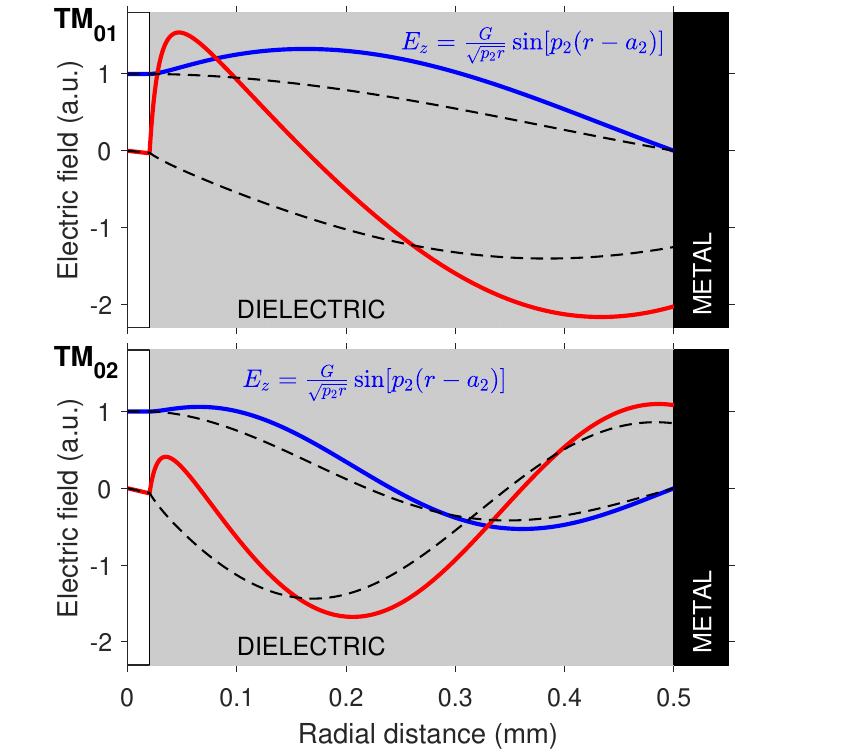}
\caption{Same as Fig.~\ref{fig:sine1} but with a smaller vacuum core radius ($20 \mum$).}
\label{fig:sine2}
\end{figure}

The approximate operating wavelengths determined using the sinusoid method are within $5\%$ of the real operating wavelengths, and the method is very accurate in the case shown in Fig.~\ref{fig:sine1}. Higher-order modes have shorter operating wavelengths, which translates to larger transverse wavenumber $p_2$ in the dielectric layer. Thus, the argument inside the Bessel functions is larger, and the asymptotic sinusoid approximation becomes better for higher-order TM modes, which is visible in the figure. For certain cases, the Bessel function arguments can get too small for the sinusoid approximation to work well. Figure~\ref{fig:sine2} is the same as {\color{black}Fig.~}\ref{fig:sine1}, but now the vacuum core is now only $20 \mum$ in radius. The sinusoidal approximation leads to considerably worse predictions for the mode field profiles.

The breakdown of the sinusoid approximation is evident in Fig.~\ref{fig:sine2}. Interestingly, the relative errors in the operating wavelengths predicted by the sinusoid method are smaller than Fig.~\ref{fig:sine2} might suggest: $\sim 26\%$ for TM$_{01}$ and $\sim 9\%$ for TM$_{02}$. What is noteworthy about the breakdown of the sinusoid method, though, is that it incorrectly predicts a positive $r$-slope for $E_z$ in the dielectric at the vacuum boundary, causing $E_z$ to seemingly peak in the dielectric. In reality, the slope is always negative, and for example the TM$_{01}$ mode should have $E_z$ falling monotonically from unity to zero with increasing $r$ inside the dielectric layer.

The TM$_{02}$ cases in Figs.~\ref{fig:sine1} and \ref{fig:sine2} are shown just to illustrate how the sinusoid method works better for higher-order modes, and from now on we will focus solely on TM$_{01}$. Since TM$_{01}$ is the lowest-order TM mode, it is characterized by the property that $E_z$ stays positive in the dielectric and only reaches zero at the metal cladding. Thus, when $a_1 < r < a_2$, the argument of the sine in Eq.~\ref{eq:sine2} must be between $-\pi$ and $0$, or otherwise $E_z$ would have a zero strictly inside the dielectric. In fact, the argument should be expected to be between $\sim -\pi/2$ and $0$ in the dielectric to correctly account for the fact that TM$_{01}$ should monotonically decrease from unity to zero inside the dielectric. Since the argument of the sine term in Eq.~(\ref{eq:sine2}) is strictly between $-\pi$ and $0$, the whole expression including the square root term can be well approximated by its third-order Taylor series in $r$ evaluated at $r = a_2$. The Taylor series expansion is
\begin{align}
&G \sqrt{ \frac{2}{\pi p_2 r} } \sin \left( p_2 (r - a_2) \right) \\
\approx - &G \sqrt{ \frac{2}{\pi} } \frac{ \sqrt{ p_2 a_2 } }{24 a_3} [r - a_2] 
\left[ 4 p_2^2 a_2^2 (r - a_2)^2  - 45 a_2^2 + 30 a_2 r - 9 r^2 \right] \\
=   &C [r - a_2] \left[ 4 p_2^2 a_2^2 (r - a_2)^2  - 45 a_2^2 + 30 a_2 r - 9 r^2 \right]  \label{eq:Tay1} ,
\end{align}
where all the constants have been absorbed into a new constant $C$. The factor $r - a_2$ can trivially be seen to take care of the metal cladding boundary condition ($E_z = 0$ at $r = a_2$). With the normalization $A_1 = 1$, i.e. $E_z = 1$ in the vacuum core, we get
\be
 C = \left\{ [a_1 - a_2] \left[ 4 p_2^2 a_2^2 (a_1 - a_2)^2  - 45 a_2^2 + 30 a_2 a_1 - 9 a_1^2 \right] \right\}^{-1} .
\ee
With this choice of $C$, all the boundary conditions of $E_z$ have been accounted for. The only boundary condition left is then that of $E_r$, i.e., that $n^2 E_r$ be continuous at the vacuum-dielectric interface. In the core we have $E_r(r) = - k_0 r / 2$ and in the dielectric we can get $E_r$ from $E_z$ through
\begin{align}
	E_r(r) &= \frac{\beta}{p_2^2} \frac{\pd E_z(r)}{\pd r} \nonumber \\ 
	&= C \frac{k_0}{p_2^2} \left[ (12 p_2^2 a_2^2 - 27)(r - a_2)^2 - 48 a_2^2 + 24 a_2 r \right] ,
\end{align}
where we have used $\beta = k_0$ for the propagation constant since the effective index is unity. Keeping in mind that the constant $C$ is dependent on $p_2$ but without diving into the full algebraic details, the matching of $n^2 E_r$ across the vacuum-dielectric interface yields a $4^\text{th}$ order polynomial equation in $p_2$. The polynomial equation only contains even orders of $p_2$, so $p_2^2$ can be solved using the quadratic formula. Knowing $p_2^2$, the physically feasible solution for $p_2$ can then be gotten by requiring $p_2 >0$. The solution is
\be
	p_2 = \sqrt{ \frac{ s_1 - \sqrt{3(s_2 + s_3 + s_4)} }{8 a_1 a_2^2 (a_2 - a_1)^2} } , \label{eq:p}
\ee
where
\begin{align}
	s_1 &= 45 a_1 a_2^2 - 30 a_1^2 a_2 + 9 a_1^3 + 24 a_2^2 n^2 (a_2 - a_1)                     \label{eq:s1} \\
	s_2 &= (27 a_1^4 - 180 a_1^3 a_2 + 570 a_1^2 a_2^2 - 900 a_1 a_2^3 + 675 a_2^4) a_1^2      \label{eq:s2}    \\
	s_3 &= 16 ( 27 a_1 a_2^2 + 9 a_1^3 - 31 a_1^2 a_2 - 5 a_2^3 ) a_1 a_2^2 n^2                  \label{eq:s3}    \\
	s_4 &= 192 a_2^4 (a_2 - a_1)^2 n^4 .                                                         \label{eq:s4}  
\end{align}
The expression for $p_2$, Eq.~(\ref{eq:p}), though lengthy, contains nothing but the fiber parameters $a_1$, $a_2$, and $n$, and gives the approximate transverse wavenumber $p_2$ in the dielectric, from which the approximate operating wavelength $\lambda_0$ can be obtained through
\be
	\lambda_0 = \frac{2 \pi \sqrt{n^2 - 1}}{p_2} \approx 
	2 \pi \sqrt{  \frac{8 (n^2 - 1) a_1 a_2^2 (a_2 - a_1)^2}{ s_1 - \sqrt{3(s_2 + s_3 + s_4)} } } .
	\label{eq:lambdaop}
\ee

\begin{figure}[h!]
\centering\includegraphics[width=\columnwidth]{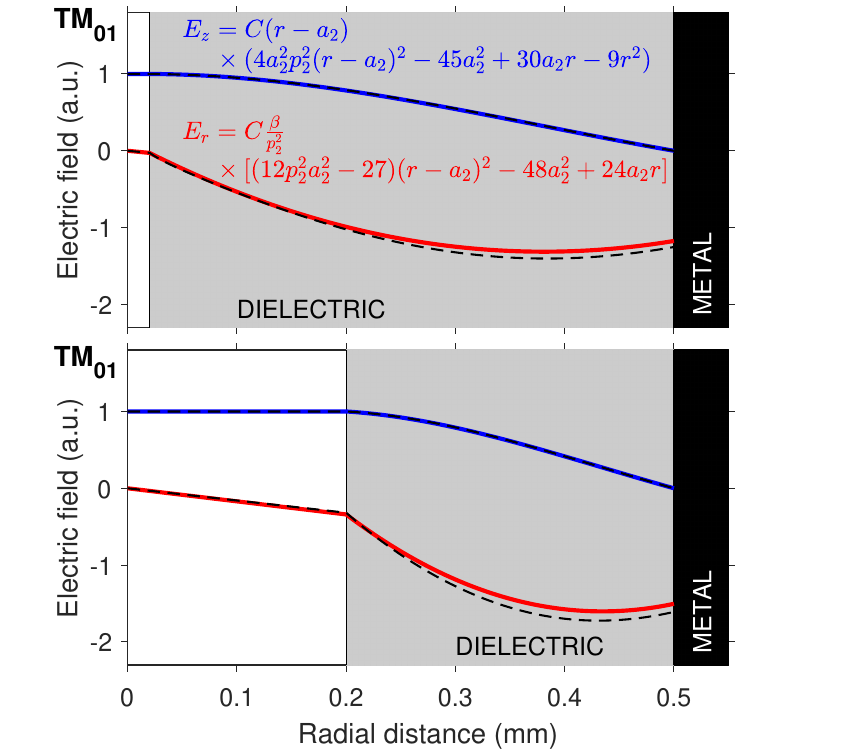}
\caption{Same as the top panels of Figs.~\ref{fig:sine1} (bottom) and \ref{fig:sine2} (top) but using the polynomial approximation instead of the sinusoidal approximations. The parameter $C$ appearing in the expressions is the scaling constant that normalizes $E_z$ to unity in the core. The red curves show $n^2 E_r$, which is the continuous quantity, but the expression given in the figure is for $E_r$.}
\label{fig:pol1}
\end{figure}

To get an understanding how accurate this approximate method is, Fig.~\ref{fig:pol1} replicates the top panels of Figs.~\ref{fig:sine1} and \ref{fig:sine2} but using the polymonial sine-Taylor method. Somewhat surprisingly, the sine-Taylor method, which is essentially an approximation upon an approximation, works better for small cores than the sinusoidal method, as can be seen in Fig.~\ref{fig:pol1}.

\begin{figure}[h!]
\centering\includegraphics[width=\columnwidth]{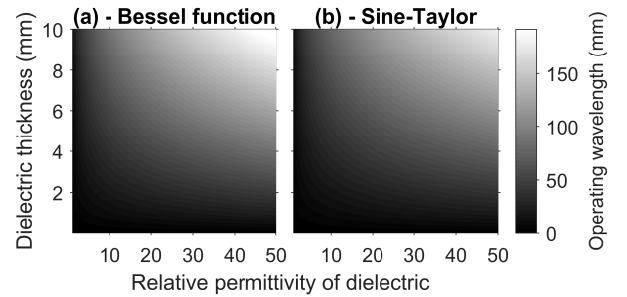}
\caption{The operating wavelengths of metal-clad hollow-core fibers with one dielectric layer computed using the true Bessel function method (a), and the sine-Taylor method (b). Note that the full range of parameters studied is not shown, but the behavior outside the range shown is the same, just with different numerical values for the operating wavelength.}
\label{fig:lambda}
\end{figure}

The relative errors in the operating wavelength predictions using the sine-Taylor method for the fibers in Fig.~\ref{fig:pol1} are $5.1\%$ for the $20\mum$ vacuum core and $6.6\%$ for the $200\mum$ core, underlining the accuracy of the method. Importantly, the polynomial approximation fully captures the behavior of the operating wavelength when the core diamater, dielectric thickness, or the dielectric permittivity are changed. This was verified numerically by fixing the core radius to $500\mum$ and determining the real operating wavelength using the Bessel function treatment and the approximate operating wavelength using the sine-Taylor method for permittivity values between $1.05$ and $200$ and dielectric layer thicknesses between $5\mum$ and $50000\mum$. The true operating wavelengths as well as the predictions from the sine-Taylor method are shown in Fig.~\ref{fig:lambda}.

\begin{figure}[h!]
\centering\includegraphics[width=\columnwidth]{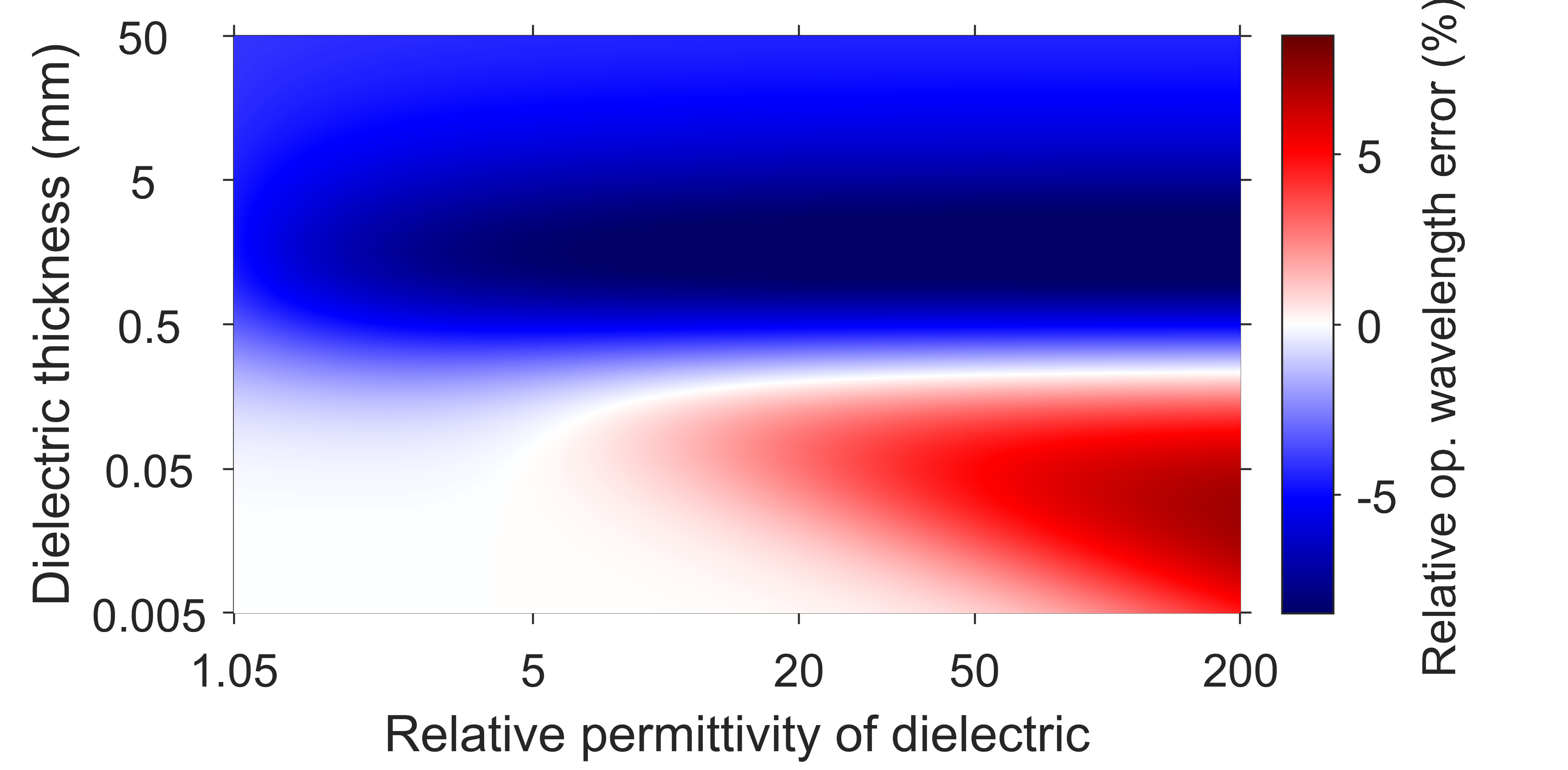}
\caption{The relative error of the sine-Taylor method in determining the operating wavelength as a function of dielectric layer thickness and permittivity. Note that both the horizontal as well as the vertical scales are logarithmic for better visualization of the behavior of the relative error.}
\label{fig:error}
\end{figure}

It is evident from Fig.~\ref{fig:lambda} that the sine-Taylor method accurately captures the behavior of the operating wavelength with respect to the dielectric layer thickness and permittivity, as the differences between Figs.~\ref{fig:lambda}(a) and \ref{fig:lambda}(b) are hard to see. To characterize the accuracy of the sine-Taylor method, we computed the relative error defined as $(\lambda_\text{approximate} - \lambda_\text{true} )/\lambda_\text{true}$. The relative error is shown in Fig.~\ref{fig:error}. The error in the predicted operating wavelength increases somewhat with the permittivity of the dielectric, but all throughout the relative error was less than $8.5\%$ for this vast range of parameters. Again, due to the scale invariance of Maxwell's equations, these results apply to any fiber where the dielectric permittivity is between $1.05$ and $200$ and the dielectric layer thickness is between $0.01$ and $100$ times the vacuum core radius.

We now have a reasonably accurate closed-form expression for the approximate operating wavelength in Eq.~(\ref{eq:lambdaop}) together with Eqs.~(\ref{eq:s1}-\ref{eq:s4}). Utilizing this expression is considerably faster in determining the operating wavelength of a fiber compared to the proper Bessel function treatment, which could be useful in fiber design for waveguide-based particle acceleration. However, the main benefits of the sine-Taylor method lie in its analytic nature. First, it explicitly shows how the operating wavelength behaves when fiber parameters are changed, which is not at all evident from the Bessel function treatment. Second, the closed form expression for the operating wavelength now makes it possible to derive approximate conditions for the dispersion of the dielectric in the fiber to have both phase and group velocity of the TM$_{01}$ mode be equal to the speed of light in vacuum. These conditions and limitations are discussed in the next section.

\section{Dielectric Requirements to Have a TM$_{01}$ Mode in a Hollow-Core Fiber with $c$ Phase and Group Velocity}

For dispersive materials, the refractive index depends on the wavelength (or equivalently, on the vacuum wavenumber). In section 3 it was shown that it is impossible to have the group velocity equal to $c$ at the operating wavelength, for which the phase velocity is $c$, using non-dispersive materials. This is not the case if we allow the refractive index $n$ of the dielectric to have an arbitrary wavelength-dependence. In this section we will outline the steps to derive the material dispersion requirements to have unity group index at the operating wavelength.

Let $k_0$ be the vacuum wavenumber corresponding to the operating wavelength. Once again we introduce a small perturbation $k_0 \rightarrow (1 + \delta) k_0$. With this perturbation, the refractive index $n(k)$, now a function of vacuum wavenumber $k$, changes as
\be
	n(k_0 + \delta k_0) = n(k_0) + \delta k_0 \frac{d n}{d k} = n(k_0) + \delta k_0 n'(k_0)
\ee
to first order, where the derivative of $n(k)$ is evaluated at the operating wavenumber $k_0$ and denoted by $n'(k_0)$. Since we want to keep the effective index at unity in the presence of the perturbation $\delta k_0$, the approximate perturbed transverse wavenumber $p_2'$ in the dielectric layer is still given by Eq.~(\ref{eq:p}) but with $k_0$ replaced by $(1 + \delta) k_0$ and $n(k_0)$ replaced by $n(k_0) + \delta k_0 n'(k_0)$. On the other hand, the perturbed transverse wavenumber is also given by
\be
	p_2' = (1 + \delta) k_0 \sqrt{ [n(k_0) + \delta k_0 n'(k_0)]^2 - 1} .
\ee
Equating these two expressions, expanding them as Taylor series in $\delta$, and only retaining terms that are zeroth and first-order in $\delta$ gives an equation from which $n'(k_0)$ can be solved. The solution reads
\begin{align}
	&n'(k_0) =  \label{eq:dndk} \\ 
	 - &k_0 (n^2 - 1) \left[ k_0^2 n - \frac{3 n}{a_1 (a_2 - a_1)} + \frac{\sqrt{3}(s_3 + 2 s_4)}{16 n_2 a_1 a_2^2 (a_2 - a_1)^2 s_5 } \right]^{-1}, \nonumber
\end{align}
where we have denoted $n(k_0)$ by $n$, $s_5 = \sqrt{3 a_1^2 (3 a_1^2 - 10 a_1 a_2 + 15 a_2^2)^2 + s_3 + s_4  }$, and $s_{3,4}$ are given by Eqs.~(\ref{eq:s3} - \ref{eq:s4}). $k_0$ is obtained from $k_0 = p_2 / \sqrt{n^2 - 1}$ where the unperturbed transverse wavenumber $p_2$ is given by Eq.~(\ref{eq:p}) using the unperturbed $n = n(k_0)$ for the refractive index. Naturally, the expression for $n'(k_0)$ given in Eq.~(\ref{eq:dndk}) is lengthy, but again it only involves the fiber parameters $a_1$, $a_2$, and $n$. Equation~(\ref{eq:dndk}) thus gives the approximate condition for the dielectric dispersion in order to have unity group index at the operating wavelength.

If me multiply both sides of Eq.~(\ref{eq:dndk}) by $k_0$ and write $a_2 = (1 + x) a_1$, the right hand side of Eq.~(\ref{eq:dndk}) simplifies to a rational function of $x$ that monotonically rises from $n - n^3$ at $x = 0$ (no dielectric layer) to $-n + 1/n$ as $x \rightarrow \infty$ (dielectric thickness goes to infinity). The inequality $-n + 1/n > n - n^3$ holds for all $n > 1$, and therefore we can write an approximate but fundamental condition for the material dispersion as
\be
	k_0 n'(k_0) < -n(k_0) + \frac{1}{n(k_0)} . \label{eq:fund}
\ee
Equation~(\ref{eq:fund}) shows that the required dispersion is generally heavily anomalous. For example, for $a_1 = 200 \mum$, $a_2 = 500 \mum$, $n = 2.1$ (quartz in the THz range), Eq.~(\ref{eq:dndk}) yields a slope of 
\be
	\frac{d n(\lambda)}{d \lambda} \approx 9.72 \cdot 10^{-7} \frac{1}{\text{nm}} = 0.972 \frac{1}{\text{mm}}
\ee
at the operating wavelength of $1.975$~mm, clearly corresponding to extreme anomalous dispersion. Given the accuracy of the sine-Taylor method used in deriving this numerical value, it can be expected to be close to the actual dispersion required.

As dictated by Kramers-Kronig relations, such anomalous dispersion can only occur within a loss window of a dielectric, and the significant increase in $n$ with wavelength is associated with heavy losses. Kramers-Kronig relations together with Eq.~(\ref{eq:fund}) thus show that any ordinary dielectric that would allow for unity group index at the operating wavelength is so lossy that the whole scheme becomes infeasible. For the example above, using $\varepsilon = n^2 = 4.41$ for the dielectric again and assuming a Lorentzian absorption spectrum at the operating wavelength with a spectral full width at half-maximum of $5$~GHz, the $1/e$ intensity attenuation length (skin depth) of the dielectric would be a mere quarter of the wavelength. The TM mode would thus experience extremely heavy losses upon propagation and the pulse would essentially disappear before any meaningful distance, making the fiber useless for relativistic electron acceleration.

To overcome this otherwise fundamental limitation, we propose using gain media for the dielectric in the fiber. The Hilbert transform of a constant function is zero, and thus Kramer's-Kronig relations are agnostic to the base level of the real and imaginary parts of the susceptibility. Therefore, the real part of the susceptibility $\chi(\omega)$ would behave similarly near an absorption peak as it would near a dip in gain. A refractive index that increases with wavelength in the presence of gain has been experimentally demonstrated and used for superluminal light propagation in rubidium gas \cite{Steinberg1994}, for example. Other media can be used as well, and manipulating the dispersive properties of materials with various forms of gain is well-understood \cite{Boyd2009}. Generally, the required dip in gain has been achieved through dual-color pumping of a medium with a narrow Raman spectrum \cite{Wang2000}, but for example some plastics readily have the required double peak in their Raman spectra \cite{Ferraro1961}. We also note that even though the analysis above was performed for a fiber with a single dielectric layer, more layers can be utilized, which would allow for the use of gases as part of the fiber design by trapping them between the metal cladding and a solid inner dielectric layer. In the next section, we will consider such a fiber and show that it is theoretically possible to achieve $c$ phase and group velocity in the THz regime in it.

\section{Hollow-Core Gain-Dip Fiber}

Wang, Kuzmich, and Dogariu demonstrated transparent anomalous dispersion in cesium vapor through the use of a Raman gain doublet \cite{Wang2000, Dogariu2001}. The probe wavelength was $894$~nm, and the experimentally demonstrated (locally linear) change in the refractive index due to the gain doublet was $\Delta n = -1.8 \cdot 10^{-6}$ over a range of $1.9$~MHz. Since the gain was based on Raman scattering, the optical frequency experiencing gain and anomalous dispersion can be changed by pumping at a different wavelength. Furthermore, the magnitude of anomalous dispersion can be changed by tuning the gain, which in turn is linearly proportional to the Raman pump power. At a frequency of $0.6$~THz, the demonstrated decrease of $\Delta n = -1.8 \cdot 10^{-6}$ over $1.9$~MHz would correspond to a $k$-slope of
\be
	k_0 n'(k_0) = -0.5684 .
\ee
Using the theoretical maximum of $1.31$ for the refractive index of atomic vapors \cite{Keaveney2012}, we have $-0.5684 < -1.31 + 1/1.31 \approx -0.5466$, and hence trivially $k_0 n'(k_0) < -n_\text{vapor} + 1/n_\text{vapor}$ for all $1 < n_\text{vapor} \leq 1.31$. The fundamental inequality of Eq.~(\ref{eq:fund}) is thus satisfied for the experimentally demonstrated scheme in \cite{Wang2000}, and therefore the vapor with its gain-modified dispersion profile would be a suitable medium to get the group index to unity for a $0.6$~THz pulse in a hollow-core fiber. However, the vapor would need to be separated from the vacuum core by a solid material.

\begin{figure}[h!]
\centering\includegraphics[width=\columnwidth]{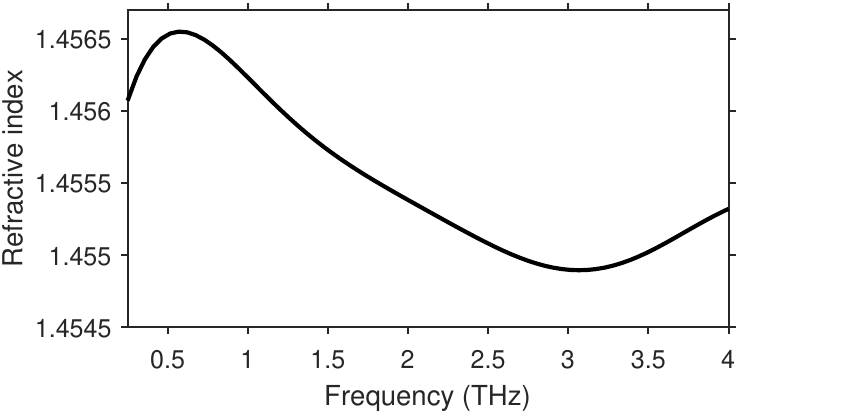}
\caption{The refractive index of polymethylpentene in the THz regime (after \cite{Podzorov2008}).}
\label{fig:TPX}
\end{figure}

Polymethylpentene, commonly called by its trademarked name TPX, is a plastic with excellent optical, chemical, and mechanical properties. It is also robust in the presence of temperature fluctuations and transparent in the THz regime. Its refractive index is shown in Fig.~\ref{fig:TPX}.

\begin{figure}[h!]
\centering\includegraphics[width=\columnwidth]{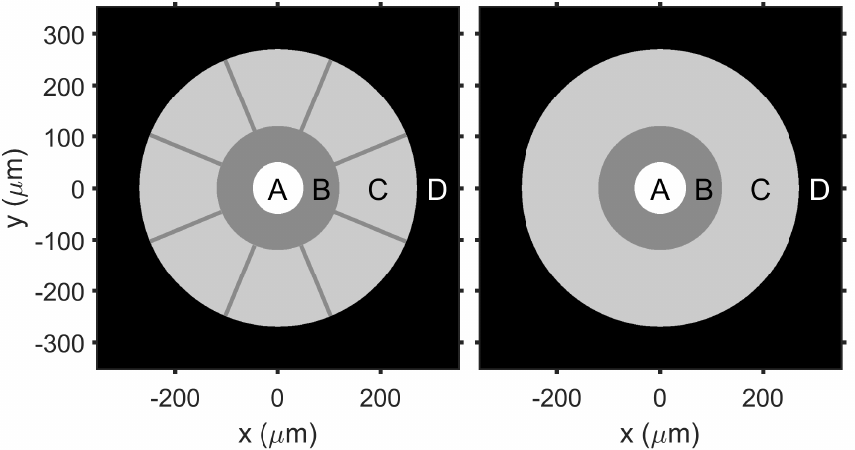}
\caption{The proposed fiber design (a) and the fiber used to model it (b). The layers from the core to the cladding are vacuum (A), polymethylpentene (B), Cs vapor (C), metal (D). The outer radii of the dielectric layers are $50\mum$, $120\mum$, and $270\mum$, respectively. The metal cladding is assumed to be a perfect conductor and thick enough (mm-scale) such that it can be assumed infinite. The refractive index of the Cs vapor is taken to be $1.2$.}
\label{fig:fiber}
\end{figure}

Consider a fiber with a vacuum core of $50\mum$ radius surrounded by a polymethylpentene layer that is $70\mum$ thick, consistent with the thickness of commercially available polymethylpentene thin films. Let the polymethylpentene layer be surrounded by a $150$-$\mu$m-thick layer of Cs vapor, the refractive index of which we (pessimistically) assume to be $1.2$ without any applied gain. The proposed fiber design is shown in Fig.~\ref{fig:fiber}(a), and it includes struts to hold the polymethylpentene layer in place. Strut thicknesses of $50$~nm have been experimentally demonstrated in lead silicate glass \cite{EbendorffHeidepriem2009}, for example, so the struts in the proposed fiber could reasonably be made thin enough so that the fiber can be accurately modeled like the one shown in Fig.~\ref{fig:fiber}(b) \cite{Ji2018}, especially since the operating wavelength is in the regime of hundreds of micrometers. Also note that Fig.~\ref{fig:fiber}(a) shows eight struts, but less could be used.

\begin{figure}[h!]
\centering\includegraphics[width=\columnwidth]{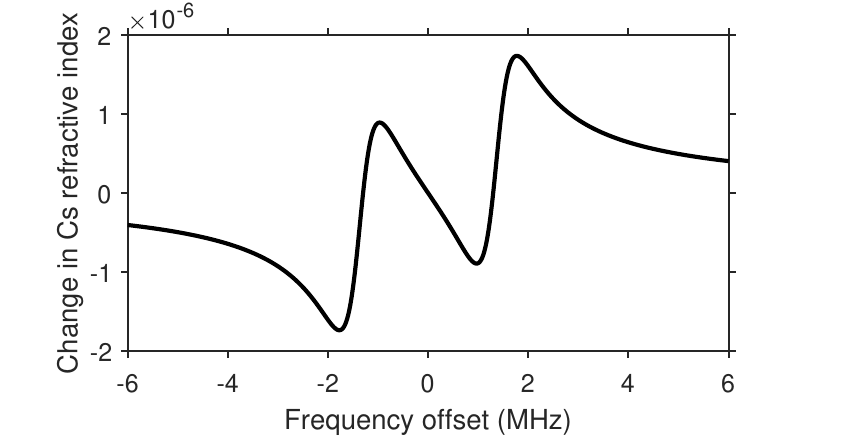}
\caption{The change in the refractive index of Cs vapor due to gain demonstrated in \cite{Dogariu2001} as a function of frequency detuning from the operating wavelength.}
\label{fig:deltan}
\end{figure}

\begin{figure}[h!]
\centering\includegraphics[width=\columnwidth]{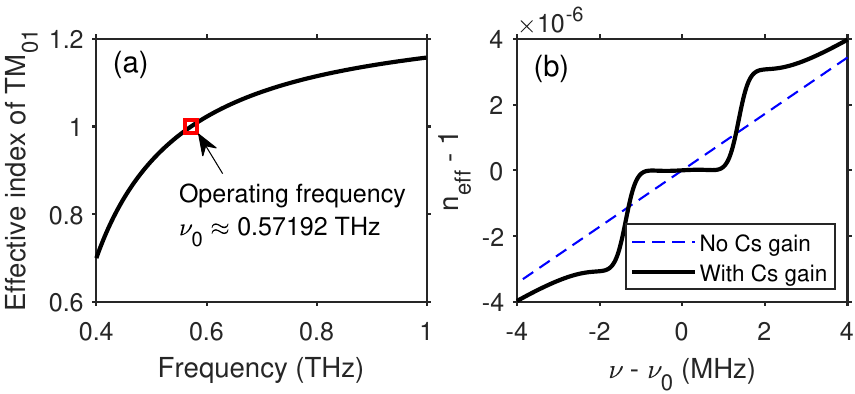}
\caption{The dispersion profile of the fiber shown in Fig.~\ref{fig:fiber}. (a) shows the dispersion profile of the TM$_{01}$ mode without gain in the Cs vapor layer. (b) is a zoom-in of the dispersion profile around the operating frequency, showing the dispersion profile with (black solid line) and without (blue dashed line) gain in the Cs layer. The gain is $94\%$ of what it was in the experimental demonstration in \cite{Dogariu2001}. The introduction of the gain makes the dispersion profile locally flat around the operating wavelength, and both the phase and group velocity of the TM$_{01}$ mode at the operating frequency are now equal to the speed of light in vacuum.}
\label{fig:dispersion}
\end{figure}

The operating wavelength of the fiber shown in Fig.~\ref{fig:fiber}, at which the phase velocity equals $c$, is approximately $0.52418$~mm, corresponding to a frequency of $\nu_0 = 0.57192$~THz. Next we add gain to the Cs layer in the same manner that was experimentally demonstrated in \cite{Dogariu2001}, centered at the operating frequency $\nu_0$. This changes the refractive index of the layer around the operating wavelength, and the refractive index change demonstrated in \cite{Dogariu2001} is very well approximated by
\be
	\Delta n = A \frac{\nu - \nu_0 - \nu_1}{1 + \left( \frac{\nu - \nu_0 - \nu_1}{\Delta \nu} \right)^2 } +
	           A \frac{\nu - \nu_0 + \nu_1}{1 + \left( \frac{\nu - \nu_0 + \nu_1}{\Delta \nu} \right)^2 }, \label{eq:deltan}
\ee
where $A = 6.5348$~THz$^{-1}$, $\nu_1 = 1.3655$~MHz, and $\Delta \nu = 0.4209$~MHz. {\color{black} Note that $\Delta \nu$ is a parameter in the expression for $\Delta n$, not a spectral width of a pulse.} The index change is shown in Fig.~\ref{fig:deltan}.
Adding the gain to the Cs layer around the operating wavelength does not alter the operating wavelength, as the gain-induced index change at the operating wavelength is zero. However, in the vicinity of the operating wavelength, the refractive index of the Cs layer changes when the gain is on, and the dispersion profile of the fiber becomes modified around the operating wavelength. Note that the amplitude $A$ in Eq.~(\ref{eq:deltan}) is dependent on the Raman pump power and can thus be tuned. By reducing the gain demonstrated in \cite{Dogariu2001} by $6\%$ for the Cs layer for the fiber proposed here, i.e. adding $0.94$ times the index change shown in Fig.~\ref{fig:deltan}, the dispersion curve can be made locally flat around the operating wavelength, which is shown in Fig.~\ref{fig:dispersion}. Figure~\ref{fig:dispersion}(a) shows the fiber's dispersion profile with the Cs layer not providing gain ($n_\text{Cs} = 1.2$), and Fig.~\ref{fig:dispersion}(b) shows the same dispersion profile zoomed around the operating frequency with and without gain in the Cs layer. 

Figure~\ref{fig:dispersion}(b) shows a locally flat dispersion profile over a $1$~MHz range around the operating wavelength. Within this range, a pulse in the TM$_{01}$ mode can thus propagate with both its phase and group velocity equal to $c$. The operating wavelength is $0.52418$~mm, which is well within the useful THz regime that has been used for electron acceleration \cite{Nanni2015}. While the sensitivity of the dispersion profile of proposed fiber to manufacturing imperfections and to the gain provided pose engineering challenges, the proposed design shows that there is nothing fundamentally preventing us from overcoming the brutal anomalous dispersion requirement of Eq.~(\ref{eq:fund}) through gain-assisted tweaking of a medium's refractive index. The tunability of the gain can also compensate for whatever minor effects the struts or the finite conductivity of the metal cladding might have on the dispersion profile, and the dispersion can always be made locally flat. The gain in the example fiber above was provided through Raman pumping, so its practicality becomes a matter of light source availability at the operating wavelength for both the probe beam accelerating the electrons and the two pumps providing the Raman gain for the probe beam. The medium of choice here was Cs vapor, but suitable solid Raman gain media would simplify the design, as the solid-vapor double layer could simply be replaced by a single gain-providing layer. The gain would not have to be Raman-based either, and any kind of gain capable of turning a material's dispersion anomalous while not compromising transparency should suffice.

The analysis here was only concerned with bringing the slope $\partial n / \partial \lambda$ to some value required to have $c$ group velocity at the operating wavelength. By even more careful gain engineering, the second derivative $\partial n / \partial \lambda$ could also be tuned, allowing for zero group velocity dispersion while simultaneously having $c$ group velocity at the operating wavelength. Since the dispersion engineering relies on gain in a fiber layer, it also has the additional advantage that the electromagnetic pulse doing the particle acceleration gets amplified upon propagation. Now the accelerating pulses do not need to be ultrashort and intense, as the length of the fiber accelerator can simply be made longer when both phase and group velocity equal $c$. Intensity is obviously still desirable, so we point out that the gain bandwidth can be increased through increasing the power of the Raman pump beams and pushing them further away from the operating wavelength in the spectral domain, which would allow for flat dispersion over a broader wavelength range and hence shorter and more intense accelerating pulses { \color{black} without significant temporal broadening due to dispersion. Instead of a dual-tone Raman pump, the gain-providing pumps could also be spectrally tailored to provide gain such that the dispersion is flat across a wider range, hindering temporal broadening.}

{ \color{black} Even when the dispersion is flat across a narrow spectral range such as above, shorter THz pulses can still be used for acceleration. Unlike the experimentally demonstrated cm-scale hollow-core accelerators \cite{Nanni2015}, the proposed fiber accelerator is not limited by group velocity mismatch. Thus, the early stage of acceleration using short pulses would be similar to but more efficient than that of existing accelerators. After the already more efficient initial stage, the phase and group velocity of the pulse remain at $c$, and hence the electrons will continue to accelerate even when the pulse is spectrally broad and suffering from dispersion outside of the flat dispersion band.

The phase velocity $c$ is dictated by the fiber dimensions and the center frequency of the accelerating pulse. It is unaffected by the gain as long as the gain does not shift spectrally, and such stability is readily achieved by wavelength-stable Raman pumps. Note, however, that the \emph{magnitude} of the gain might easily oscillate due to pump power fluctuations, which would then cause the \emph{group velocity} to vary. At first this might seem like a significant practical challenge, but such fluctuations would simply move the envelope of the pulse with respect to the electrons. Thus, if the \emph{average} level of the gain can be kept at its desired value, the pulse envelope would merely hover around the electron bunch, still providing gain on average. Given that the accelerating pulses would eventually become longer than a few cycles (either through dispersion or by virtue of having been long to begin with), the pulse envelope dancing around the electrons would not even appreciably decrease the average accelerating electric field, and absolute pump power stability is not required. The proposed scheme might thus be able to scale accelerator lengths from the centimeter scale to tens of meters or even more.}

\section{Conclusions}

We theoretically studied metal-clad hollow-core fibers with dielectric layers between the core and the cladding for relativistic charged particle acceleration applications. An accurate approximate mathematical method dubbed the sine-Taylor method was developed to determine the operating wavelength at which the phase velocity of the TM$_{01}$ mode is equal to $c$, the speed of light in vacuum. The method was then used to derive an approximate expression for the material dispersion of the dielectric layer in terms of fiber parameters required to have the group velocity also equal to $c$ at the operating wavelength. The expression showed that the dispersion would need to be so strongly anomalous for all fiber parameters and wavelengths that losses would render the fiber useless for particle acceleration. We then proposed utilizing gain in the dielectric layer to make it anomalously dispersive without a loss window. In theory, this allows for propagation and amplification of TM$_{01}$-mode electromagnetic pulses in the THz regime and elsewhere with $c$ phase and group velocity, making it possible to arbitrarily scale the length of hollow-core fiber electron accelerators and hence the energy gain of the electrons.

\begin{backmatter}

\bmsection{Disclosures}
The authors declare no conflicts of interest.

\bmsection{Contributions}
S. Ramachandran proposed the project to design a fiber with $c$ phase and group velocity, and oversaw the work. A. Antikainen developed the theory, the mathematical methods, and the fiber design example. The manuscript was prepared jointly.

\end{backmatter}

\bibliography{refs}


\end{document}